\input harvmac
\noblackbox

\input epsf

\newcount\figno
\figno=0
\def\fig#1#2#3{
\par\begingroup\parindent=0pt\leftskip=1cm\rightskip=1cm\parindent=0pt
\baselineskip=11pt
\global\advance\figno by 1
\midinsert
\epsfxsize=#3
\centerline{\epsfbox{#2}}
\vskip 12pt
{\bf Fig.\ \the\figno: } #1\par
\endinsert\endgroup\par
}
\def\figlabel#1{\xdef#1{\the\figno}}
\def\encadremath#1{\vbox{\hrule\hbox{\vrule\kern8pt\vbox{\kern8pt
\hbox{$\displaystyle #1$}\kern8pt}
\kern8pt\vrule}\hrule}}

\def\adag{a^{\dagger}}

\font\cmss=cmss10
\font\cmsss=cmss10 at 7pt
\def\rlx{\relax\leavevmode}
\def\inbar{\vrule height1.5ex width.4pt depth0pt}
\def\IN{\relax{\rm I\kern-.18em N}}
\def\IP{\relax{\rm I\kern-.18em P}}
\def\ZZ{\rlx\leavevmode\ifmmode\mathchoice{\hbox{\cmss Z\kern-.4em Z}}
 {\hbox{\cmss Z\kern-.4em Z}}{\lower.9pt\hbox{\cmsss Z\kern-.36em Z}}
 {\lower1.2pt\hbox{\cmsss Z\kern-.36em Z}}\else{\cmss Z\kern-.4em
 Z}\fi}
\def\IZ{\relax\ifmmode\mathchoice
{\hbox{\cmss Z\kern-.4em Z}}{\hbox{\cmss Z\kern-.4em Z}}
{\lower.9pt\hbox{\cmsss Z\kern-.4em Z}}
{\lower1.2pt\hbox{\cmsss Z\kern-.4em Z}}\else{\cmss Z\kern-.4em
Z}\fi}
\def\IZ{\relax\ifmmode\mathchoice
{\hbox{\cmss Z\kern-.4em Z}}{\hbox{\cmss Z\kern-.4em Z}}
{\lower.9pt\hbox{\cmsss Z\kern-.4em Z}}
{\lower1.2pt\hbox{\cmsss Z\kern-.4em Z}}\else{\cmss Z\kern-.4em
Z}\fi}

\def\narrowplus{\kern -.04truein + \kern -.03truein}
\def\narrowminus{- \kern -.04truein}
\def\narrowminussub{\kern -.02truein - \kern -.01truein}

\def\t{{\theta}}
\def\l{{\lambda}}

\def\frac#1#2{{#1\over #2}}

\def\IZ{\relax\ifmmode\mathchoice
{\hbox{\cmss Z\kern-.4em Z}}{\hbox{\cmss Z\kern-.4em Z}}
{\lower.9pt\hbox{\cmsss Z\kern-.4em Z}}
{\lower1.2pt\hbox{\cmsss Z\kern-.4em Z}}\else{\cmss Z\kern-.4em
Z}\fi}
\def\IC{{\relax\,\hbox{$\inbar\kern-.3em{\rm C}$}}}
\def\p{\partial}
\font\cmss=cmss10 \font\cmsss=cmss10 at 7pt
\def\IR{\relax{\rm I\kern-.18em R}}
\def\ra{\rangle}
\def\la{\langle}

\def\IZ{\relax\ifmmode\mathchoice
{\hbox{\cmss Z\kern-.4em Z}}{\hbox{\cmss Z\kern-.4em Z}}
{\lower.9pt\hbox{\cmsss Z\kern-.4em Z}}
{\lower1.2pt\hbox{\cmsss Z\kern-.4em Z}}\else{\cmss Z\kern-.4em
Z}\fi}
\def\IC{{\relax\,\hbox{$\inbar\kern-.3em{\rm C}$}}}

\def\a{{\alpha}}

\def\ep{{\epsilon}}

\def\t{{\theta}}
\def\l{{\lambda}}

\def\P{{\Phi}}

\def\t{{\theta}}

\def\r{{\rightarrow}}

\def\frac#1#2{{#1\over #2}}

\def\adag{a^{\dagger}}
\def\ri{\rho_{MM}(\l)}
\def\rj{\rho_{GT}(\l)}

\def\CN{{\cal N}}

\def\L{{\Lambda}}

\def\p{\partial}

\lref\th{G.~'tHooft, ``A Planar Diagram Theory for Strong Interactions,'' Nucl.\ Phys.\ {\bf 72},
461, (1974).}
\lref\wit{E. ~Witten, in {\it Recent Developments in Gauge Theories} eds. G 'tHooft et.al. Plenum
Press, New York and London (1980).}
\lref\mig{A. ~Migdal, Ann.\ Phys.\ {\bf 109}, 365, (1977).}
\lref\gv{
R.~Gopakumar and C.~Vafa,
``On the gauge theory/geometry correspondence,''
Adv.\ Theor.\ Math.\ Phys.\  {\bf 3}, 1415 (1999)
[arXiv:hep-th/9811131].}
\lref\ov{
H.~Ooguri and C.~Vafa,
``Worldsheet derivation of a large N duality,''
Nucl.\ Phys.\ B {\bf 641}, 3 (2002)
[arXiv:hep-th/0205297].}
\lref\malda{
J.~M.~Maldacena,
``The large $N$ limit of superconformal field theories and supergravity,''
Adv.\ Theor.\ Math.\ Phys.\  {\bf 2}, 231 (1998)
[Int.\ J.\ Theor.\ Phys.\  {\bf 38}, 1113 (1999)]
[arXiv:hep-th/9711200].}
\lref\vafa{
C.~Vafa,
``Superstrings and topological strings at large N,''
J.\ Math.\ Phys.\  {\bf 42}, 2798 (2001)
[arXiv:hep-th/0008142].}
\lref\civ{
F.~Cachazo, K.~A.~Intriligator and C.~Vafa,
``A large N duality via a geometric transition,''
Nucl.\ Phys.\ B {\bf 603}, 3 (2001)
[arXiv:hep-th/0103067].}
\lref\cv{
F.~Cachazo and C.~Vafa,
``N = 1 and N = 2 geometry from fluxes,''
arXiv:hep-th/0206017.}
\lref\dvone{
R.~Dijkgraaf and C.~Vafa,
``Matrix models, topological strings, and supersymmetric gauge theories,''
Nucl.\ Phys.\ B {\bf 644}, 3 (2002)
[arXiv:hep-th/0206255].}
\lref\dvtwo{
R.~Dijkgraaf and C.~Vafa,
``On geometry and matrix models,''
Nucl.\ Phys.\ B {\bf 644}, 21 (2002)
[arXiv:hep-th/0207106].}
\lref\dvthree{
R.~Dijkgraaf and C.~Vafa,
``A perturbative window into non-perturbative physics,''
arXiv:hep-th/0208048.}
\lref\dglvz{
R.~Dijkgraaf, M.~T.~Grisaru, C.~S.~Lam, C.~Vafa and D.~Zanon,
``Perturbative computation of glueball superpotentials,''
arXiv:hep-th/0211017.}
\lref\dgvv{
R.~Dijkgraaf, S.~Gukov, V.~A.~Kazakov and C.~Vafa,
``Perturbative analysis of gauged matrix models,''
arXiv:hep-th/0210238.}
\lref\doug{M. Douglas, ``Mastering $\CN=1$.'' Talk at ``Strings 2002'', 
Cambridge, U.K.
.}
\lref\fgz{P.Di~ Francesco, P. ~Ginsparg, J. ~Zinn-Justin,
``2D Gravity and Random Matrices,'' Phys.Rept. {\bf 254} (1995) 1.}
\lref\dorthree{
N.~Dorey, T.~J.~Hollowood and S.~P.~Kumar,
``S-duality of the Leigh-Strassler deformation via matrix models,''
arXiv:hep-th/0210239.}
\lref\acgh{
R.~Argurio, V.~L.~Campos, G.~Ferretti and R.~Heise,
arXiv:hep-th/0210291.}
\lref\nsw{S.~Naculich, H.~Schnitzer, and N.~Wyllard,
``The N=2 gauge theory prepotential and periods from a perturbative matrix model calculation,''
hep-th/0211123.}
\lref\mcg{
J.~McGreevy,
``Adding flavor to Dijkgraaf-Vafa,''
arXiv:hep-th/0211009.}
\lref\gor{
A.~Gorsky,
``Konishi anomaly and N = 1 effective superpotentials from matrix models,''
arXiv:hep-th/0210281.}
\lref\dortwo{N.~Dorey, T.~J.~Hollowood, S.~P.~Kumar and A.~Sinkovics,
``Massive vacua of N = 1* theory and S-duality from matrix models,''
hep-th/0209099.}
\lref\sw{N.~Seiberg, E.~Witten, ``Monopole Condensation,
And Confinement In N=2 Supersymmetric Yang-Mills Theory,''
Nucl.Phys. {\bf B426} (1994) 19; Erratum-ibid. {\bf B430} (1994) 485.}
\lref\Klemm{A.~Klemm, W.~Lerche and S.~Theisen,
``Nonperturbative effective actions of N=2 supersymmetric gauge theories,''
Int.\ J.\ Mod.\ Phys.\ A {\bf 11} (1996) 1929, hep-th/9505150.}
\lref\dor{N.~Dorey, T.~J.~Hollowood, S.~Prem Kumar and A.~Sinkovics,
``Exact superpotentials from matrix models,'' hep-th/0209089.}
\lref\av{M.~Aganagic and C.~Vafa,
``Perturbative derivation of mirror symmetry,'' hep-th/0209138.}
\lref\ff{F.~Ferrari,
``On exact superpotentials in confining vacua,'' hep-th/0210135.}
\lref\fo{H.~Fuji, Y.~Ookouchi,
``Comments on Effective Superpotentials via Matrix Models,''
hep-th/0210148.}
\lref\ber{D.~Berenstein, ``Quantum moduli spaces from matrix
models,'' hep-th/0210183.}
\lref\cm{L.~Chekhov and A.~Mironov,
``Matrix models vs. Seiberg-Witten/Whitham theories,'' hep-th/0209085.}
\lref\suz{H. Suzuki,''Perturbative Derivation of Exact Superpotential for 
Meson Fields from Matrix Theories with One Flavour,''
hep-th/0211052.}
\lref\br{I.~Bena, R.~Roiban, 
``Exact superpotentials in N=1 theories with flavor and their matrix model formulation,''
hep-th/0211075.}
\lref\fftwo{F. ~Ferrari, 
``Quantum parameter space and double scaling limits in N=1 super Yang-Mills theory,''
hep-th/0211069.}
\lref\dj{Y.~Demasure, R.~A.~Janik, ``Effective matter superpotentials from Wishart random matrices,''
hep-th/0211082.}
\lref\bipz{E.~Brezin, C.~Itzykson, G.~Parisi and J.~B.~Zuber,
``Planar Diagrams,'' Commun.\ Math.\ Phys.\  {\bf 59}, 35 (1978).}\
\lref\ils{
K.~A.~Intriligator, R.~G.~Leigh and N.~Seiberg,
``Exact superpotentials in four-dimensions,''
Phys.\ Rev.\ D {\bf 50}, 1092 (1994)
[arXiv:hep-th/9403198].}
\lref\ds{
M.~R.~Douglas and S.~H.~Shenker,
``Dynamics of SU(N) supersymmetric gauge theory,''
Nucl.\ Phys.\ B {\bf 447}, 271 (1995)
[arXiv:hep-th/9503163].}
\lref\spr{G.~Springer,
``Introduction to Riemann Surfaces,''
Chelsea, New York, 1981.}
\lref\cohn{H. ~Cohn,
``Conformal Mapping on Riemann Surfaces,''
Dover, New York, 1967.}
\lref\klyt{
A.~Klemm, W.~Lerche, S.~Yankielowicz and S.~Theisen,
``Simple singularities and N=2 supersymmetric Yang-Mills theory,''
Phys.\ Lett.\ B {\bf 344}, 169 (1995)
[arXiv:hep-th/9411048].}
\lref\af{
P.~C.~Argyres and A.~E.~Faraggi,
``The vacuum structure and spectrum of N=2 supersymmetric SU(n) gauge theory,''
Phys.\ Rev.\ Lett.\  {\bf 74}, 3931 (1995)
[arXiv:hep-th/9411057].}
\lref\haan{
O.~Haan,
Z.\ Phys.\ C {\bf 6}, 345 (1980).}
\lref\halsch{
M.~B.~Halpern and C.~Schwartz,
``Large N Classical Solution For The One Matrix Model,''
Phys.\ Rev.\ D {\bf 24}, 2146 (1981).}
\lref\gg{
R.~Gopakumar and D.~J.~Gross,
``Mastering the master field,''
Nucl.\ Phys.\ B {\bf 451}, 379 (1995)
[arXiv:hep-th/9411021].}
\lref\douglas{
M.~R.~Douglas,
``Stochastic master fields,''
Phys.\ Lett.\ B {\bf 344}, 117 (1995)
[arXiv:hep-th/9411025].}

\Title
{\vbox{\baselineskip12pt
\hbox{hep-th/0211100}}}
{\vbox{\centerline{${\cal N}=1$ Theories and a Geometric Master Field}}}

\centerline{Rajesh Gopakumar\foot{gopakumr@mri.ernet.in}}

\centerline{\sl Harish-Chandra Research Institute, Chhatnag Rd.,}
\centerline{\sl Jhusi, Allahabad, India 211019}
\medskip

\vskip 0.8cm

\centerline{\bf Abstract}
\medskip
\noindent

We study the large $N$ limit of the class of $U(N)$ ${\CN}=1$ SUSY gauge theories
with an adjoint scalar and a superpotential $W(\P)$. In each of the vacua of 
the quantum theory, the expectation values $\la$Tr$\Phi^p$$\ra$ are determined by a master 
matrix $\Phi_0$
with eigenvalue distribution $\rho_{GT}(\l)$. $\rho_{GT}(\l)$ is quite 
distinct from the 
eigenvalue distribution $\rho_{MM}(\l)$ of the corresponding large $N$ matrix model proposed by 
Dijkgraaf and Vafa. Nevertheless, it has a simple 
form on the auxiliary Riemann surface of the matrix model. 
Thus the underlying geometry of the matrix model 
leads to a definite prescription for 
computing $\rho_{GT}(\l)$, knowing $\rho_{MM}(\l)$.

\vskip 0.5cm
\Date{Nov. 2002}

\newsec{Introduction}

For a long while after 'tHooft pointed out the emergence of Riemann surfaces in the large N 
expansion \th , the relationship between gauge theories and geometry remained only a 
tantalising picture. However, in recent years following Maldacena's conjecture
\malda , we have begun
to understand more clearly the nature of the gauge theory/geometry correspondence. 

Topological strings provide a very tractable context in which to precisely study this correspondence. 
By now, the original duality of this kind \gv\ between large N Chern-Simons theory and closed 
topological strings has been understood at different levels (See for e.g. \ov ).   
Subsequently, embedding these topological string dualities 
in physical string theory \vafa\ has led to a lot of new insights.
For instance,
Cachazo, Intriligator and Vafa \civ\ pointed out the geometric origin of the quantum superpotential
of a large class of ${\cal N}=1$ theories in this way.   
These results and the relation to topological strings led 
Dijkgraaf and Vafa \dvone\ to the striking conjecture that these gauge theory superpotentials were 
determined by a large $N$ zero dimensional matrix model. This conjecture 
has been subsequently generalised and 
checked in various other cases \dvtwo\ \dvthree \cm\ \dor\ \dortwo\ \av\ \ff\ \fo\ \ber\
\dgvv\ \dorthree\ \gor\ \acgh\ \mcg\ \suz\ \fftwo\ \br\ \dj . 
Very recently a pertubative field theory argument for the localisation of 
these gauge theory superpotentials to zero dimensional matrix integrals, has also been given 
\dvthree\ \dglvz . 

This geometrising of our understanding of large $N$ gauge theories means that in some sense, 
the geometry is the master field \wit\ \mig\ 
of the gauge theory (as also observed in \dvthree\ \doug ). 
Moreover, the relation to the planar 
matrix model for $\CN=1$ theories holds out hope of obtaining a simple characterisation of 
the master field of these theories (at least in the holomorphic sector). In fact, one might 
guess that the large $N$ saddle point of the matrix integral would play the role of the 
master field. This is not quite the case. 
Nevertheless, 
in this note we will make a beginning in trying to understand how exactly the matrix model 
(and its underlying geometry) can play the role of the master field, at least, for the  
class of ${\cal N}=1$ theories studied in \civ . 

These $\CN=1$ $U(N)$ gauge theories, 
have an adjoint chiral superfield $\Phi$ with a tree level superpotential $W(\Phi)$ which is 
generally taken to have a polynomial form
\eqn\wtree{W(\Phi)=\sum_{p=2}^{n+1}{g_p\over p}\rm{Tr}\Phi^p.}
There are many quantum vacua of the theory 
each labelled by a specific pattern of gauge symmetry breaking. 
For a given tree level $W(\P)$, Dijkgraaf and Vafa gave a prescription for 
obtaining the effective superpotential (even for the finite rank theory)
from the eigenvalue density of the large $N$ matrix model 
with potential $W(\P)$ \dvone . 

To make progress towards identifying a master field in the matrix model (or geometry) description, 
we will also need to look at 
other holomorphic quantities of interest in the gauge theory (GT for short). In fact, 
the class of observables that are natural to examine in this context 
are the expectation values, in each of these vacua, of
$\la Tr\P^p\ra_{GT}$ (for arbitrary $p$, in the limit of large $N$) 
\foot{Note that 
the low energy superpotential $W_{low}(g_p ,\L)$ obtained from the Dijkgraaf-Vafa 
prescription gives us some
information in this regard: it computes $\la Tr\P^p\ra$ for $p\leq n+1$. 
For instance, in the case of a cubic superpotential we have ${\p W_{low}\over \p g_2}=\la Tr\P^2\ra$ 
and  ${\p W_{low}\over \p g_3}=\la Tr\P^3\ra$. Higher moments can be obtained through a deformation
of the superpotential. See \nsw .}. 
The solution of the gauge theory via deformation of the underlying $\CN=2$ theory \civ\ \cv\
contains the answer, in principle, for the values of these observables.  

Here, we will merely make the observation 
that this information is also naturally contained in the matrix model,
being encoded in the associated riemann surface.  
However, belying naive expectations, $\la Tr\P^p\ra_{GT}$
are {\it not} the same as the expectation values $\la Tr\P^p\ra_{MM}$ in the matrix model
\foot{This was also observed in the matrix model analysis of the 
confining vacuum of the $\CN=1^*$ theory \dor .}. Thus the gauge theory master field matrix $\P_0$
(which would reproduce the moments $\la Tr\P^p\ra_{GT}$) 
is not the saddle point of the large $N$ matrix integral.
Instead, $\P_0$ is characterised by a distinct eigenvalue distribution $\rj$. We can, however,
give a definite prescription for extracting $\rj$ 
from knowing the matrix eigenvalue distribution
$\ri$. We will see that despite being distinct, $\rj$ and $\ri$ have many similarities
including the fact that they have identical support over the same intervals.  

There are actually good reasons why we should not have expected $\la Tr\P^p\ra_{GT}$
to coincide with $\la Tr\P^p\ra_{MM}$. For a general \wtree\ there is always a confining 
vacuum -- where the gauge symmetry is unbroken. The low energy physics in this vacuum 
should be given by pure $\CN=1$ super Yang-Mills and this should be 
reflected in $\la Tr\P^p\ra_{GT}$ (and thus $\rj$)
being independent of the detailed form of the superpotential $W(\P)$.
But in the matrix model, $\ri$ depends very strongly on the form of $W(\P)$. Thus, clearly 
the two cannot coincide in general. We will see explicitly how our prescription will 
nonetheless enable us to extract out the universal $\rj$ out of the nonuniversal $\ri$ for 
these cases.   

In the next section we review the solution to the $\CN=1$ gauge theory and the bare bones of 
the Dijkgraaf-Vafa prescription. In section 3, we proceed from this solution to show how 
the matrix model knows about the density of eigenvalues $\rj$ of the gauge theory. We give a 
calculationally precise prescription for extracting this from the matrix model solution. 
In Section 4 we illustrate this prescription with a couple of simple examples.

\newsec{Review}

In this section we review known results of \civ\ \cv\ \dvone\
keeping more or less to the original notation of 
those papers.

The theory with a tree level superpotential \wtree\ has classical vacua 
preserving $\CN=1$ SUSY when the eigenvalues of $\P$ take 
values in the set of zeroes $\{a_i\}$ (assumed for simplicity to be distinct) 
of $W'(x)=g_{n+1}\prod_{i=1}^n(x-a_i)$. For a $U(N)$ gauge theory, the vacua can then 
be labelled by the positive integers $N_i$ corresponding to the number of eigenvalues taking
the value $a_i$ (with $\sum N_i=N$). Classically the gauge symmetry is broken 
\eqn\gauge{U(N)\r\prod_{i=1}^nU(N_i).}
Quantum mechanically, only the $n$ $U(1)$'s in the product gauge group survive in the low energy 
theory since the $SU(N_i)$ sectors are confining. There is a low energy effective superpotential
$W_{low}(g_p,\L)$ corresponding to each of these vacua, where $\L$ is the dynamically generated 
scale of the underlying $SU(N)$ theory. Sometimes, it is convenient to think in terms of a 
Veneziano-Yankielowicz type effective superpotential 
$W_{eff}(S_i, g_p, \L_i)$ 
which depends on glueball superfields $S_i=Tr_{SU(N_i)}W_{\a}W^{\a}$ and scales $\L_i$. 
$W_{low}$ is then obtained 
by minimising $W_{eff}$ w.r.t. the $S_i$ and then evaluating it on the minimum.  

We can extract the information about holomorphic quantities in the quantum vacua (for instance,
$W_{low}$)
from the Seiberg-Witten (SW) curve \sw\
\eqn\swc{\tilde{y}^2=P_N^2(x)-4\L^{2N}}
of the corresponding $U(N)$ $\CN=2$ Yang-Mills theory \klyt\ \af . 
This is because the vacua of the 
$\CN=1$ theory are points on the coulomb branch of the $\CN=2$ theory \sw . In the spirit of
\ds\ Cachazo,
Intriligator and Vafa \civ\ observed that the symmetry breaking in \gauge\ implies that
the $\CN=1$ vacua would lie on submanifolds of the coulomb branch where the SW curve factorises
as 
\eqn\swfact{\tilde{y}^2=P_N^2(x)-4\L^{2N}=F_{2n}(x)H_{N-n}^2(x)}
where $F_{2n}$ and $H_{N-n}$ represent polynomials of degree $2n$ and $N-n$ respectively. 
This factorisation with $N-n$ double roots reflects $N-n$ (mutually local) 
monopoles becoming massless. The $\CN=1$ vacua are the points on this submanifold where 
the tree level superpotential would be minimised.  
Further, as was proven by Cachazo and Vafa \cv , these minima are uniquely determined 
\foot{We also require that in the $\L\r 0$ limit, $P_N(x)\r \prod_{i=1}^n(x-a_i)^{N_i}$.} 
by requiring that
$F_{2n}(x)$ be of the form 
\eqn\fw{g_{n+1}^2F_{2n}=W'(x)^2+f_{n-1}(x),}
i.e. it is a deformation of the classical ($\L\r 0$) answer by a polynomial of degree $n-1$.
The effect of this deformation is to split the double roots $a_i$ of $W'(x)^2$ into (generically)
distinct pairs of roots $(a_i^-,a_i^+)$.

The crucial observation \civ\ (motivated by the large $N$ dual Calabi-Yau geometry)
then is that the reduced Riemann surface specified by 
\eqn\redsw{y^2=F_{2n}(x)}
completely determines $W_{eff}$ and thus $W_{low}$ (as also the $U(1)$ gauge couplings). 
The genus $n-1$ hyperelliptic riemann surface in \redsw\ has $n$ branch cuts $\a_i$
between the $n$ pairs of roots $(a_i^-,a_i^+)$ of $F_{2n}$. Then $W_{eff}$ is given by 
\eqn\period{S_i={g_{n+1}\over 2\pi i}\oint_{\a_i}y dx ; ~~~~~~~~ 
\Pi_i={g_{n+1}\over 2\pi i}\int_{C_i}y dx; 
~~~~~~~~ W_{eff}=\sum_i(N_i\Pi_i+\tau_{YM}S_i),}
where the curves $C_i$ start from the $i$th branch cut and go off to infinity 
in the $x$ plane (see Fig.1) of \cv . 
Unlike the original SW curve which has a strong dependence on $N$, the curve \redsw\
is universal in the sense that it depends only on the pattern of symmetry breaking as specified 
by the ratios $\nu_i={N_i\over N}$.

More precisely, consider $U(K)\r \prod_{i=1}^n U(K_i)$ with $\sum_i  K_i=K$, with the $K_i$ 
having no common divisor. We are then localised to points on the coulomb branch where the SW 
curve satisfies
\eqn\uk{P_K^2 -4\L^{2K}=F_{2n}H_{K-n}^2(x)}
with $F_{2n}$ as in \fw .
Now consider the $U(N)$ theory (with $N=MK$) with the same classical superpotential 
$W(\P)$ and the same 
pattern of symmetry breaking, namely, $U(N)\r \prod_{i=1}^n U(N_i)$, where $N_i=MK_i$.
It is easy then to check that 
\eqn\sws{P_{(N=MK)}(x)=\tilde{\L}^{MK}T_M({P_K(x)\over \L^k})}
satisfies 
\eqn\newsw{P_{MK}^2-4\tilde{\L}^{2MK}=F_{2n}H_{MK-n}^2(x).}
where 
\eqn\hn{H_{MK-n}(x)={\tilde{\L}^{MK}\over\L^k}U_{M-1}({P_K(x)\over \L^k})H_{K-n}(x).} 
Here $T_M(z=2\cos\t)=2\cos M\t$ and $U_{M-1}(z=2\cos\t)={\sin M\t\over \sin\t}$ are 
the usual Chebyshev polynomials.

The point here is that the SW curve of the 
$U(MK)$ theory has the same factorisation as the $U(K)$ theory with the 
{\it same} $F_{2n}$. Thus we associate the same riemann surface \redsw\ to all these theories.
This riemann surface can only depend on the fractions $\nu_i$. 
In fact, this is why the geometry is able to capture
even the finite $K$ superpotential. 
But we can also take 
$M\r \infty$ and thus take the large $N$ limit while preserving the pattern 
of symmetry breaking. $W_{eff}$, and after minimisation $W_{low}$, are determined by \period\ 
in terms of the riemann surface \redsw .

Dijkgraaf and Vafa \dvone\
observed that \redsw\ is nothing other than the riemann surface associated to the 
large $N$ limit of the zero dimensional hermitian matrix model 
\eqn\mm{Z_{MM}=\exp(-{N^2\over \mu^2}F_0)=\int [D\P]\exp(-{N\over \mu}TrW(\P)),} 
expanded about the vacuum with a fractional 
pattern of symmetry breaking (not necessarily related to that of the gauge theory). 
This connection was also motivated by the fact that the superpotential of this gauge theory
is captured by an open topological string theory which reduces to this matrix model.

The prescription in 
the matrix model to compute the superpotential is then as follows.
From the solution of the matrix model, 
the saddle point eigenvalue distribution $\ri$ for \mm\ can be seen to be 
proportional to the discontinuity of the 
meromorphic one form $y dx$ across the branch cuts $\a_i$. Therefore $\oint_{\a_i}y dx$
now has the matrix model interpretation of being 
proportional to the fraction of eigenvalues supported on the cut $\a_i$.
Comparing with \period\  we need the identification 
\eqn\ni{{g_{n+1}\over 2\pi i}\oint_{\a_i}y dx= S_i.} 
Given $S_i$ this equation can be 
used to determine the coefficents of $f_{n-1}$ that appears in $F_{2n}$. The planar free energy 
$F_0$ expanded about the vacuum of the matrix model with this distribution of eigenvalues then obeys
\eqn\free{{\p F_0\over \p S_i}={g_{n+1}\over 2\pi i}\int_{C_i}y dx =\Pi_i.}
We can then compute $W_{eff}$ using the last relation in \period\ and thus $W_{low}$ after 
minimisation w.r.t. the $S_i$. Our interest, in what follows,
will be in the particular Riemann surface underlying the 
matrix model after $S_i$ have taken their minimum values $\la S_i \ra$.

\newsec{The Master Matrix $\P_0$}

We first obtain the master matrix $\P_0$ in the gauge theory by putting 
together the various ingredients of the solution of the previous section. 
We will then see that it is given in terms of a 
particularly simple form on the Riemann surface \redsw . 
Thus we will be able to give a prescription of how how to obtain the eigenvalue distribution of 
$\P_0$ from the matrix model. Though  
distinct from $\ri$ it is nevertheless completely determined in terms of $\ri$. 

Our considerations will be for a fixed (but arbitrary) tree level $W(\P)$ as in \wtree .
Let's start with a $U(K)\r \prod_{i=1}^n U(K_i)$ vacuum.
Since by definition $P_K(x)=\la det(x-\P)\ra$, 
the $K$ roots of $P_K$ determine $\la Tr\P^p\ra_{GT}$ for $p\leq K$.
We can study the properties of the vacuum (with the same pattern of symmetry breaking)
in the large $N=MK$ limit
by scaling all the $K_i$ by a common factor $M$ (as in the previous section) and 
taking $M\r \infty$.

Since, by \sws , $P_{MK}(x)=\tilde{\L}^{MK}T_M({P_K(x)\over \L^k})$, the $(N=MK)$ roots of $P_{MK}(x)$
are given by 
\eqn\roots{P_K(\l^{(k)}_m)=2\L^K\cos{(2m+1)\pi\over 2M}, ~~~~~~~~ (m=0\ldots M-1, k=1\ldots K).} 
Here the right hand side comes from the $M$ roots of the Chebyshev polynomial $T_M$.
Taking $M\r\infty$, we see from the r.h.s that there is a uniform distribution on the semicircle 
$\t \in [0,\pi]$. 
It is easy then to verify that the distribution of the roots $\l^{k}_m$ is given by 
\eqn\dist{\rj\equiv{1\over \pi}{d\t\over d\l}={1\over \pi K}{dP_K(\l)\over d\l}
{1\over \sqrt{4\L^{2K}-P_K^2(\l)}}}
such that 
\eqn\mom{{1\over N}\la Tr\P^p\ra_{GT}=\int \l^p \rj d\l.}
Thus, from the solution to the gauge theory we see that the master matrix $\P_0$ has an eigenvalue
distribution $\rj$ given by \dist . 
This distribution has also appeared in \cv\ the context of the geometric dual to the 
$\CN=2$ Seiberg-Witten theory. 

We would now like to see if we can give a prescription to directly obtain 
this distribution from the matrix model 
and the associated reduced riemann surface \redsw . From the form \dist\ it is not 
obvious that we can do so. After all the matrix model does not know about $K$. It is only sensitive
to the filling fractions $\nu_i={K_i\over K}$.  

However, notice that, 
\eqn\uka{P_K^2 -4\L^{2K}=F_{2n}H_{K-n}^2(x)} 
implies that
\eqn\pdp{P_K{dP_K(x)\over dx}=H_{K-n}(x)[2H^{\prime}_{K-n}F_{2n}+H_{K-n}F^{\prime}_{2n}]\equiv 
H_{K-n}Q_{K+n-1}}
where $\prime$ denotes differentiation with respect to $x$. Since $P_K$ has no roots in common
with $H_{K-n}$ (that would contradict \uka ), \pdp\ implies that $Q_{K+n-1}(x)$ has $P_K(x)$ 
as a factor
$$Q_{K+n-1}(x)=P_K(x)R^{(K)}_{n-1}(x)$$
where we have defined a degree $n-1$ polynomial $R^{(K)}_{n-1}(x)$ and  
the superscript is a reminder that it can still depend on $K$ even if its degree does not.
Therefore,
\eqn\rn{{dP_K(x)\over dx}=H_{K-n}(x)R^{(K)}_{n-1}(x)}
and hence \dist\ simplifies (using \uka\ and \rn) to 
\eqn\distr{\rj={1\over \pi K}{dP_K(\l)\over d\l}{1\over \sqrt{4\L^{2K}-P_K^2(\l)}}={1\over \pi K}
{R^{(K)}_{n-1}(\l)\over \sqrt{-F_{2n}(\l)}}.}

We will now see that 
$$R_{n-1}(\l)={1\over K}R^{(K)}_{n-1}(\l)$$
is actually independent of the overall 
size $K$ of the gauge group, 
i.e. ${1\over K}R^{(K)}_{n-1}(\l)={1\over MK}R^{(MK)}_{n-1}(\l)$ for any $M$.
Then $R_{n-1}(\l)$ would depend 
only on the pattern of symmetry breaking represented by 
the ratios $\nu_i$. 

This result follows from \sws\ which after differentiating
gives 
\eqn\pnk{\eqalign{{dP_{MK}(x)\over dx}=&{\tilde{\L}^{MK}\over \L^K}MU_{M-1}({P_K(\l)\over \L^k})
{dP_K(\l)\over d\l} \cr
=&{\tilde{\L}^{MK}\over \L^K}MU_{M-1}({P_K(\l)\over \L^k})H_{K-n}(\l)R^{(K)}_{n-1}(\l) \cr
=&MH_{MK-n}(x)R^{(K)}_{n-1}(\l).}}
In obtaining the first line we have used the property of Chebyshev 
polynomials that ${dT_M(x)\over dx}=MU_{M-1}(x)$. 
The next two lines follow from 
Eqs. \rn\ and \hn . 
Comparing the last line with Eq.\rn\ for $K$ replaced by $MK$, namely,
$${dP_{MK}(\l)\over d\l}=H_{MK-n}(\l)R^{(MK)}_{n-1}(\l),$$
we see that
\eqn\rmk{{1\over MK}R^{(MK)}_{n-1}(\l)={1\over K}R^{(K)}_{n-1}(\l).}
Thus proving the claim of the previous paragraph.

As a result, we see that \distr\ can be written in the universal form 
\eqn\distri{\rj={1\over \pi}{R_{n-1}(\l)\over \sqrt{-F_{2n}(\l)}}.}
As mentioned in Sec.2 ,
the eigenvalue value density $\ri$ of the matrix model with potential $W(\P)$ 
expanded about a vacuum with filling fractions $\la S_i\ra$ is determined by the one form $ydx$ as 
in \ni\ to be  
\eqn\rmm{{\mu\over g_{n+1}}\ri={1\over 2\pi}\sqrt{-F_{2n}(\l)}.}
Thus the two distributions 
are quite distinct, though we see that they have the same cut structure and 
thus identical support over the $n$ branch cuts of the riemann surface \redsw . Classically, the 
eigenvalues of the matrix model sit at the extrema $a_i$ of the potential $W(\P)$. As do the 
eigenvalues of $\P$ in the gauge theory. We see that quantum mechanically, the eigenvalues
of both spread over the intervals $[a_i^-,a_i^+]$ 
defined by the branch cuts of $\sqrt{-F_{2n}}$. 
Thus the $\ri$ and $\rj$ are temptingly similar but nonetheless distinct.  
And for good reason, too, as we argued in the introduction.

Can we, knowing the large $N$ solution $\ri$ to the matrix model, reconstruct $\rj$? The answer is 
yes, 
essentially because $\rj$ can be expressed in  terms of a natural one form on the riemann surface 
determined by the matrix model solution. 
In fact, $\rj$ can be written in terms of the meromorphic one form 
\eqn\mer{\omega={R_{n-1}(x)dx \over y}}
on the riemann surface $y^2=F_{2n}(x)$. The discontinuity of $\omega$ across the cuts
is proportional to $\rj$. In fact,
\eqn\peri{{1\over 2\pi i}\oint_{\a_i}\omega=\int_{a_i^-}^{a_i^+}\rj d\l =\nu_i.}   
The last equality follows from the fact that the fraction of gauge theory eigenvalues in the $i$th 
vacuum is $\nu_i$. It can also be seen to follow from the form of $\rj$ in \dist .
This fact will enable us to construct $\omega$ knowing $ydx$ 
or equivalently $\ri$.

Note that the $n-1$ one forms 
$${x^{k}dx\over y}, ~~~~~~~~ (k=0\ldots n-2)$$
are holomorphic on our riemann surface and form a basis for the 
space of holomorphic one forms (for this and other facts 
about hyperelliptic surfaces used below, see for example
\spr\ or \cohn\ ). 
On the other hand, ${x^{n-1}dx\over y}$
has a simple pole at each of the two preimages of $x=\infty$ (with residue $\pm 1$). Thus $\omega$
is a one form which has only (two) simple poles coming 
from the highest power of $x$ in $R_{n-1}(x)$. From the definition of $R_{n-1}$ (see \rn\ and below) 
and since 
the coefficent of the 
highest power of $x$ in $P_K$, as is that of $F_{2n}$ (from Eq. \fw ), we conclude that 
the coefficent of $x^{n-1}$ in $R_{n-1}(x)$ is 
also one. Therefore $\omega$ has residues $\pm 1$ at its poles. 

Note then that the one form 
$$\omega^{\prime}=\omega-{x^{n-1}dx\over y}$$
is a holomorphic 
one form by construction. We know its A-periods, since we know the periods of $\omega$ and we 
can calculate the periods of ${x^{n-1}dx\over y}$. Knowing the A-periods  
uniquely determines a holomorphic form -- we have a unique expansion in the basis of 
holomorphic forms given above. We can thus construct $\omega^{\prime}$ and therefore 
$\omega$ and $\rj$ 
uniquely, once we know the riemann surface from the solution to the matrix model.

\newsec{Examples}

It might help if the general considerations of the previous section were illustrated with a couple 
of examples. 

\subsec{Gaussian Model}

The simplest case to study is a $U(N)$ theory with quadratic superpotential
$$W(\P)={m\over 2}Tr \P^2$$
There is only one vacuum, classically at $\P=0$ and with unbroken gauge symmetry.  

The corresponding large $N$ gaussian matrix model 
\eqn\qmm{Z =\int [D\P]\exp(-{Nm\over 2\mu}Tr\P^2).}
has the well known Wigner semicircular distribution
\eqn\wig{\ri={1\over 2\pi\L^2}\sqrt{4\L^2-\l^2}, ~~~~~ \l\in[-2\L, 2\L], ~~~~~~~(\L^2={\mu\over m}).}
The planar moments are given by  
\eqn\momm{{1\over N}\la Tr\P^{2k}\ra_{MM}={(2k)!\over k!(k+1)!}\L^{2k}.}

Let's apply the prescription of the previous section to obtain $\rj$. From \wig\
and \ni\ it follows that $y^2=x^2-4\L^2$. Since $n=1$, we have from \mer\ and \peri\ that 
\eqn\rgauge{\omega={dx\over y}\Rightarrow \rj={1\over \pi}{1\over \sqrt{4\L^2-\l^2}},
~~~~~~~\l\in[-2\L, 2\L].}
$\rj$ gives rise to the distinct moments
\eqn\momgt{{1\over N}\la Tr\P^{2k}\ra_{GT}={(2k)!\over (k!)^2}\L^{2k}.}
This distribution of eigenvalues \rgauge\ in the gauge theory goes back to the work of \ds .
It arises from the factorisation of the SW curve in this vacuum as \ds\
\eqn\swq{P_N(x)=\L^NT_N({x\over \L})\Rightarrow P_N^2(x)-4\L^{2N}=U_{N-1}^2({x\over \L})(x^2-4\L^2).}
We see that in both the gauge theory and the matrix model, that the eigenvalues which were 
classically at zero are now spread over the interval $[-2\L, 2\L]$.

As an aside, we point out a curious connection of the two distributions in \wig\ and \rgauge\
with earlier work on the master field. One representation of the master field for the 
gaussian matrix model is in terms of the Cuntz oscillators \haan\ \halsch\ \gg\ \douglas\
$a$ and $\adag$ obeying
$a\adag=1$, $\adag a=1-|0\ra\la 0|$ with $a|0\ra =0$. Then
\eqn\cuntz{\hat{M}=\L(a+\adag)\Rightarrow \la 0|\hat{M}^{2k}|0\ra={1\over 2\pi\L^2}\int \l^{2k}
\sqrt{4\L^2-\l^2}d\l.}
In other words, $\hat{M}$ above is the master field for the gaussian model (see for example \gg\ 
for more details). 
What we would like to point out here is that 
\eqn\cgt{Tr\hat{M}^{2k}={1\over \pi}\int \l^{2k}{1\over \sqrt{4\L^2-\l^2}}d\l .} 
In other words, the operator trace (as opposed to cuntz vacuum expectation values)
of powers of $\hat{M}$ reproduces the gauge theory moments and in this sense is the master 
field for the gauge theory as well. The significance of this is not clear.
The fact that $\hat{M}$ has the same eigenvalue distribution 
as $\rj$, in this case, is justified in the appendix.

\subsec{Other Confining Vacua}

For any arbitrary superpotential \wtree\ we always have a vacuum with classically unbroken gauge 
symmetry. For simplicity, we will take $W(\P)$ to be even.

In the matrix model 
\eqn\gmm{Z=\int [D\P]\exp(-{N\over \mu}TrW(\P)),} 
the confining vacuum corresponds to expanding around the classical vacuum where all the eigenvalues 
are at the origin. The corresponding matrix eigenvalue density $\ri$ therefore has only one
cut and takes the form (see for e.g. \fgz\ ) 
\eqn\gr{\ri={1\over 2\pi}P_{n-1}(\l)\sqrt{4\L^2-\l^2},~~~~~ \l\in[-2\L, 2\L].} 
The actual polynomial $P_{n-1}(x)$ depends on the potential $W(\P)$ \fgz . 
This can be thought of as a special case of the riemann surface \redsw\ in which $n-1$ pairs  
of the branchpoints $(a_i^-,a_i^+)$ have coalesced. In other words 
\eqn\yconf{F_{2n}(x) \propto P_{n-1}^2(x)(x^2-4\L^2).} 
The corresponding planar 
moments will depend very much on $P_{n-1}(\l)$ and hence the details of the
potential.

However, as we explained in the introduction, 
we expect the gauge theory answers for this vacuum to be independent of the detailed 
form of the potential. We see this from our prescription in the following way.
From \yconf\ and \mer\ it follows that 
\eqn\omconf{\omega ={R_{n-1}(x)dx \over cP_{n-1}(x)\sqrt{x^2-4\L^2}},}
where $c$ is a constant.
We saw earlier that the only poles of $\omega$ are at infinity. This continues to be true when
the riemann surface degenerates due to the coalescing branchpoints. (The period integrals \peri\ over 
the other $n-1$ branch cuts are zero in this case since all the eigenvalues are spread 
about the origin. After they coalesce, these vanishing period integrals around the 
erstwhile branchpoints points imply that they are regular points, not poles.) Therefore, in \omconf\
the potential poles from the zeroes of $P_{n-1}$ must cancel against the zeroes of $R_{n-1}$. In 
other words, for this vacuum, 
$$R_{n-1}(x)= cP_{n-1}(x).$$

This implies that  
\eqn\rjconf{\rj={1\over \pi}{1\over \sqrt{4\L^2-\l^2}},}
as for the gaussian. The moments in the gauge theory also continue to be given by \momgt .
Thus our prescription enables us to extract the universal behaviour of the confining vacua out of 
the highly non-universal behaviour of the matrix model. 

\newsec{Final Comments}

We have made a start in this paper to precisely characterise the notion of a 
geometric master field at least for a class of $\CN=1$ theories. While we gave a mathematically 
precise prescription in the matrix model geometry, it would be nice to have a physical 
explication of the relation between $\ri$ and $\rj$. In this context we would like
to point out that using the matrix model prescription for extracting $\rj$, we can compute
the low energy superpotential very simply:
\eqn\wlow{W_{low}(g_p, \L)=\sum_{p=2}^{n+1}{g_p\over p}\la Tr\P^p\ra_{GT}=
N\oint_C\rj W(\l)d\l ,} 
where  
$C$ is a contour encircling all the cuts. This is different from the Dijkgraaf-Vafa
prescription (reviewed in Sec.2) for $W_{eff}$ which after minimisation gives 
\eqn\dvwlow{W_{low}(g_p,\L)=W_{eff}(g_p,\la S_i\ra, \L)
=\sum_iN_i{\p F_0\over \p S_i}\Big{|}_{\la S_i\ra}
=N\sum_i\nu_i\int{\p \over \p S_i}(\mu\ri)\Big{|}_{\la S_i\ra} W(\l)d\l.}
Here we have used the definition of $F_0$ in \mm ; we also  
recall from \ni\ that $\mu=\sum_iS_i$. Note that the term in $W_{eff}$ linear in $S_i$, 
does not contribute to $W_{low}$ after minimisation as it cancels against a nonperturbative
contribution to the free energy, from the volume of the unbroken gauge group.  
Comparing with \wlow\ suggests that 
\eqn\rirj{\rj=\sum_i\nu_i{\p \over \p S_i}(\mu\ri)\Big{|}_{\la S_i\ra}.}

In fact, after the first version of this paper appeared, a relation was presented in \nsw\ 
between gauge theory vevs and those of the matrix model, which is equivalent to \rirj .
The argument in \nsw\ (attributed to Vafa), relies on deforming the matrix model action 
\mm\ by an infinitesimal
perturbation ${\ep\over k}Tr\P^k$ and looking at the effect on $W_{eff}$.

We can also see from \rirj\ that $\rj$ takes the form \distri .
As in \cv\ we simply have to change variables while minimising $W_{eff}$.   
We take the derivatives in \rirj\ w.r.t. to the variables $b_k$ instead of $S_i$
\eqn\rirjb{\rj=\sum_{i,k}\nu_i{\p b_k\over \p S_i}{\p \over \p b_k}(\mu\ri).} 
where we parametrise the polynomial $f_{n-1}$ in \fw\ as
$$f_{n-1}(x)=\sum_{k=0}^{n-1}b_kx^k.$$
From \rmm ,\fw\ and \ni\ we then see that 
\eqn\rjf{\rj=i\sum_{i,k}\nu_i{{\l^k\over \sqrt{-F_{2n}(\l)}}
\Big{(}\oint_{\a_i}{x^k\over y}dx}\Big{)}^{-1}.}
and hence 
$\rj$ in \rirj\ is of the form \distri . It would be nice if these different forms for $\rj$
had a direct physical interpretation in the matrix model.

Though the master field as a concept is intrinsic to the large $N$ limit, it is curious that in
these $\CN=1$ theories, many large $N$ results go over to finite $N$. There should be some 
systematic way of understanding this by thinking of the large $N$ limit as a classical limit.
It will also be interesting to understand the special points in parameter space of these theories
where the conventional large $N$ limit breaks down (see \fftwo\ for a recent discussion 
of possible double scaling limits).    

The role of the Cuntz oscillators 
in Sec. 4.1 is intriguing, but might well be an accident particular to the gaussian case.
More insight into all the above questions
will also probably be had by generalising to other $\CN=1$ systems.

\newsec{Acknowledgements}

I would like to thank I. Biswas, R. Dey, D. Ghoshal, D. Jatkar and A. Sen for
very helpful conversations. I would also like to thank S. Naculich and C. Vafa for their comments 
on an earlier version.  
This work has been entirely supported by the generosity
of the people of India.

\newsec{Appendix}

Here we will show that the Cuntz operator $\hat{M}=\L(a+\adag)$ has a distribution of eigenvalues 
\eqn\rc{\rho_{C}(\l)={1\over \pi}{1\over \sqrt{4\L^2-\l^2}}.}

We can construct finite $K\times K$ matrices $a_K$, $\adag_K$
which approximate to the Cuntz oscillators $a, \adag$.
\eqn\am{(a_K)_{ij} =\delta_{i,j-1}; ~~~~~~(\adag_K)_{ij}=\delta_{i-1,j}, ~~~~~~~ (i,j=0\ldots K-1).}
$a_K$ (and its adjoint $\adag_K$) are just shift matrices with nonzero entries just above (below)
the diagonal. Note that these are different from the usual 't Hooft shift matrices which have a 
nonzero corner entry which makes them unitary.  
Then $a_K|0\ra=0$ and also
\eqn\fincuntz{a_K\adag_K={\bf I}-|K-1\ra\la K-1| , ~~~~~~~~~~~~\adag_K a_K={\bf I}-|0\ra\la 0|,}
where ${\bf I}$ is the $K\times K$ identity matrix. Our matrix indices $0\ldots K-1$ reflect 
the notation for the kets which are just the canonical unit vectors in the K-dimensional 
vector space on which these matrices act.  
These approximate, as $K\r \infty$, the Cuntz algebra $a\adag=1$, 
$\adag a=1-|0\ra\la 0|$ together with $a|0\ra =0$.  

It is not difficult to verify that $\hat{M}_K=\L(a_K+\adag_K)$ has eigenvalues 
\eqn\meig{\l_k=2\L\cos{\pi k\over K+1}, ~~~~~~~~~~~~(k=1\ldots K).}
In the large $K$ limit this leads to
the distribution \rc .

\listrefs

\end